\documentclass[a4paper]{article}

\usepackage{INTERSPEECH2019}

\usepackage{relsize}
\usepackage{bbm}
\usepackage{subcaption}

\usepackage{xcolor}
\usepackage{url}
\def\vec#1{\ensuremath{\bm{{#1}}}}

\newcommand{\R}{\ensuremath{\mathbb{R}}}

\newcommand{\x}{\ensuremath{\mathbf{x}}}

\newcommand{\seqit}[1]{\bar{#1}}

\newcommand{\sx}{\seqit{\x}}

\newcommand{\figref}[1]{Fig.~\ref{#1}}

 \newfont{\msym}{msbm10}

\newcommand{\pbij}{p_{\mathrm{b}_{i,j}}}
\newcommand{\pci}{p_{\mathrm{c}_i}}

\newcommand{\ignore}[1]{}

\makeatletter
\newcommand{\printfnsymbol}[1]{%
  \textsuperscript{\@fnsymbol{#1}}%
}
\makeatother

\title{SpeechYOLO: Detection and Localization of Speech Objects}
\name{Yael Segal\printfnsymbol{1}\thanks{\printfnsymbol{1}These authors contributed equally to this work}, Tzeviya Sylvia Fuchs\printfnsymbol{1}, Joseph Keshet}
\address{Bar-Ilan University, Ramat Gan, Israel}
\email{\{segalya, fuchstz, jkeshet\}@cs.biu.ac.il}

\begin{document}

\maketitle
\begin{abstract}
In this paper, we propose to apply object detection methods from the vision domain on the speech recognition domain, by treating audio fragments as objects. More specifically, we present SpeechYOLO, which is inspired by the YOLO algorithm \cite{redmon2016you} for object detection in images. The goal of SpeechYOLO is to localize boundaries of utterances within the input signal, and to correctly classify them. Our system is composed of a convolutional neural network, with a simple least-mean-squares loss function. We evaluated the system on several keyword spotting tasks, that include corpora of read speech and spontaneous speech. Our system compares favorably with other algorithms trained for both localization and classification.
\end{abstract}



\section{Introduction}
\label{seq:introduction}

Recently automatic speech recognition (ASR) has became ubiquitous in many applications. While ASR systems like \emph{DeepSpeech 2} \cite{amodei2016deep} and \emph{wav2letter} \cite {collobert2016wav2letter} reached amazing results in transcribing read and conversational speech, sometimes it is desired to spot and locate a predefined small set of words with extremely high accuracy. For example, services like \emph{Google Now} or \emph{Apple's Siri} can be activated by pronouncing ``OK Google'' or ``Hey Siri'', respectively \cite{chen2014small, sainath2015convolutional}. It is also used by intelligence services to accurately find specific keywords while monitoring suspected phone calls. The task of detecting and localizing words can be used to automatically analyze the diadochokinetic articulatory task \cite{fletcher1972time, westbury1993articulatory} that is used to analyze pathological speech and hence cannot be performed effectively with ASR systems. In this work we present an end-to-end system that goes from a speech signal to the transcription and alignment of given keywords (this is in contrast to the spoken term detection task that makes predictions on keywords that it has not been trained on). 

Our architecture performs both detection and localization of these predefined keywords. Previous works typically focus on only one of these two challenges. Namely, algorithms would either predict what words appear in a given utterance, thus performing detection \cite{chen2014small, sainath2015convolutional}, or are given the audio signal and the target transcription and align them, thus performing localization mostly using forced alignment \cite{keshet2007large, mcauliffe2017montreal}. Keshet \emph{et al.} \cite{keshet2009discriminative} proposed to use the confidence of a phoneme aligner and an exhaustive search to detect and localize terms that are given by their phonetic content.

In the vision domain, object detection algorithms combine the two aforementioned tasks: detection of the desired object and its localization in the image. Specifically, the YOLO \cite{redmon2016you} and SSD \cite{liu2016ssd} algorithms identify objects in an image using bounding boxes. Inspired by the idea of using  bounding boxes for object detection in images, we propose to identify speech objects in an audio signal. More specifically, consider the word classification task as a form of object detection for a speech signal. 

Palaz \emph{et al.} \cite{palaz2016jointly} presented work that is most similar to ours. Their algorithm was trained to jointly locate and classify words. However, they used a weakly supervised setting and did not use word alignments, and hence were unable to perfectly predict the whole time-span of the predicted words. In our work, however, our goal is both to detect and to locate the entire span of every word, so both tasks' results are strongly accurate.

This paper is organized as follows. In Section \ref{seq:problem_setting} we formally introduce the classification and localization problem setting. We  present our proposed method in Section \ref{seq:model}, and in Section \ref{sec:experiments} we show experimental results and  various applications of our derived method. Finally, concluding remarks and future directions are discussed in Section \ref{seq:conclusions}.

\section{Problem Setting}
\label{seq:problem_setting}

The input to our system is a speech utterance. The input speech utterance is represented as a series of acoustic feature vectors. Formally, let $\sx = (\x_1,\ldots,\x_T)$ denote the input speech utterance of a fixed duration $T$, where each $\x_t\in\R^D$ is a $D$-dimensional vector for $0\leq t \leq T$. We further define the lexicon $\mathcal{L} = \{k_1, k_2, ..., k_L \}$ to be the target set of $L$ keywords or terms that may appear in the audio signal $\sx$. Note that the utterance does not necessarily contain any of these keywords or may contain several of them. In our setting the speech objects are the $L$ keywords, but the model proposed here is not limited to specific keywords and can be used to detect and localize any audio or speech object, e.g., the syllables \emph{/pa/, /ta/,} and \emph{/ka/} in the diadochokinetic articulatory task. Our goal is to spot all the occurrences of the keywords in a given utterance $\sx$ and estimate their corresponding locations. 

We assume that $N$ keywords were pronounced in the utterance $\sx$, where $N\ge 0$. Each of these $N$ events is defined by its lexical content and its time location. Each such event $e$ is defined formally by the the tuple $e=(k,t^k_\mathrm{start},t^k_\mathrm{end})$, where $k\in\mathcal{L}$ is the actual keyword that was pronounced, and $t^k_\mathrm{start}$ and $t^k_\mathrm{end}$ are its start and end times, respectively. Our goal is therefore to find all the events in an utterance, so that for each event the correct object $k$ is identified as well as its beginning and end times.


\section{Model}
\label{seq:model}

\begin{figure}[t]
 \centering
 \includegraphics[width=\linewidth]{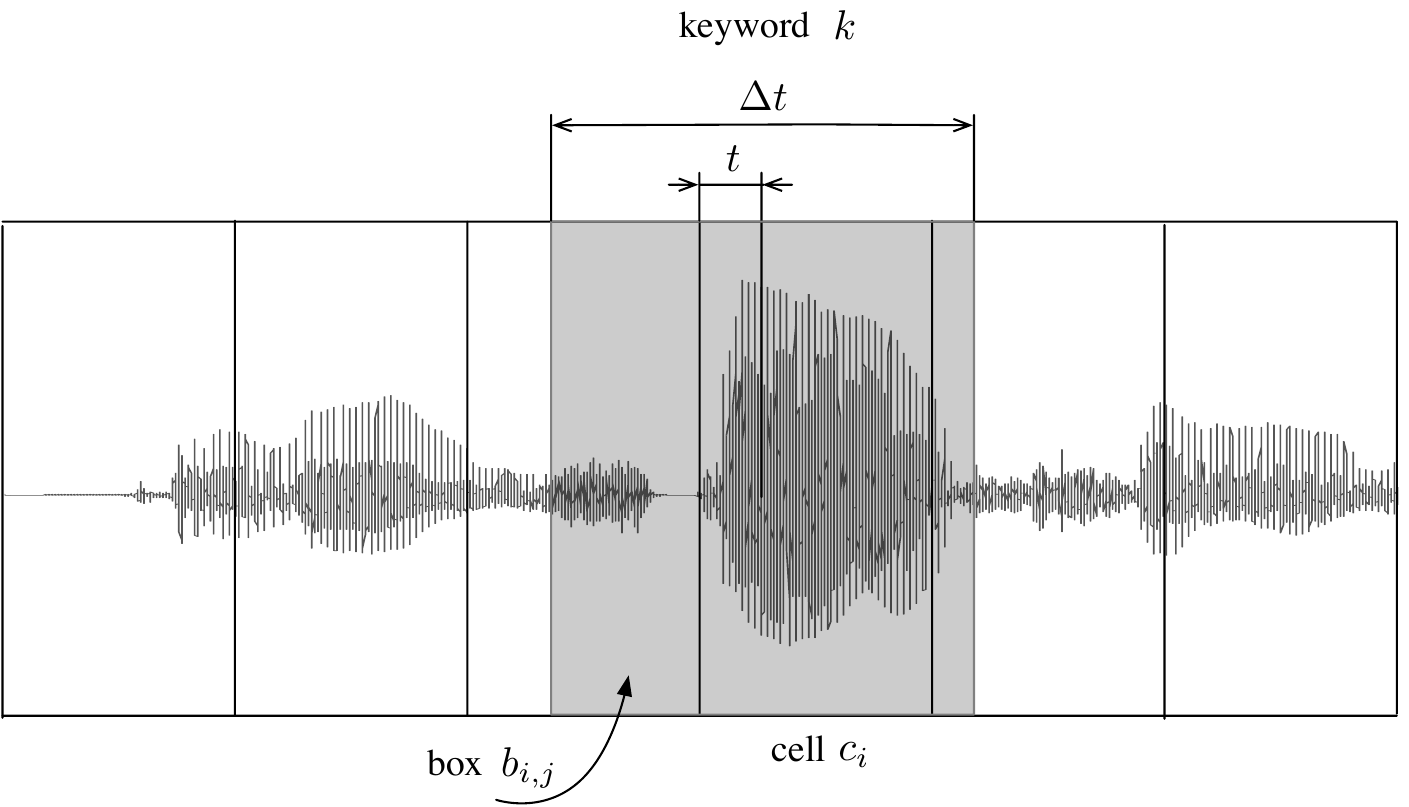}
 \caption{The notation used in our paper. The keyword ``star'' is found within cell $c_i$. One of the timing boxes $b_{i,j}$ is depicted with a shaded box, and it defines the timing of the keyword relative to the cell's boundaries.}
 \label{fig:speech_yolo_notation}
\end{figure}

As previously mentioned, our model is inspired by the YOLO model  \cite{redmon2016you}. We now describe our model formally. Our notation is schematically depicted in \figref{fig:speech_yolo_notation}. We assume that the input utterance $\sx$ is of a fixed size $T$ (1 second in our setting). We divide the input-time to $C$ non-overlapping equal sections called \emph{cells} ($C=6$ in our setting). Each cell is in charge of detecting a single event (at most) in its time-span. That is, the $i$-th cell, denoted $c_i$, is in charge of the portion $[t_{c_i}, t_{c_{i+1}}-1]$, where $t_{c_i}$ is the start-time of the cell and $t_{c_{i+1}}-1$ is its end-time, for $1\le i \le C$. The cell estimates the probability $\Pr(k | c_i)$ of each keyword $k\in\mathcal{L}$ to be uttered within its time-span. We denote the estimation of this probability by $\pci(k)$.

The cell is also in charge of localizing the detected event. The localization is defined relative to the cell's boundaries. Specifically, the location of the event is defined by the time $t \in [t_{c_i}, t_{c_{i+1}}-1]$, which is the center of the event relative to the cell's boundaries, and $\Delta t$, the  duration of the event. Note that $\Delta t$ can be longer than the time-span of the cell. Using this notation the event spans $[t_{c_i} + t-\Delta t/2 , t_{c_i} + t+\Delta t/2 ]$. 

In order to localize effectively, each cell is associated with $B$ \emph{timing boxes} (called \emph{bounding boxes} in the YOLO literature). Each box $b_{i,j}$ of the cell $c_i$ tries to independently localize the event and estimate the probability of the timing given the presumed keyword, $\Pr(t, \Delta t|k)$. It is defined by the tuple $(t_j, \Delta t_j, \pbij)$, where $\pbij$ is the confidence score of the localization $t, \Delta t$ and it can be considered as an estimation of the probability $\Pr(t, \Delta t|k, c_i)$. 

We now turn to describe the model's inference. The inference for each cell is performed independently. For the $i$-th cell, $c_i$, the chosen event is composed of the keyword $k'$ and the timing $t', \Delta t'$ that maximizes the conditional probabilities:
\begin{align} 
(k', t', \Delta t') &= \arg\max_{k, t, \Delta t} \Pr(k, t, \Delta t|c_i)\\ \label{eq:sec_eq}
&= \arg\max_{k, t, \Delta t} \Pr(k|c_i)\,\Pr(\mathrm{t, \Delta t}|k, c_i).
\end{align}
The first conditional probability in Eq.~(\ref{eq:sec_eq}) is $\pci(k)$, whereas the second conditional probability is $\pbij$ of box $b_{i,j}$. Since there are $L$ keywords and $B$ boxes the search space reduces to $L\times B$ elements, hence it is very efficient:
\[
 \max_{k \in \mathcal{L}} ~\max_{1\le j\le B} ~~~\pci(k) \,\, \pbij.
\]
Finally, the event is considered to exist in the cell if its conditional probability from above is greater than a  threshold, $\theta$.


We conclude this section by describing the training procedure. Our model, \emph{SpeechYOLO}, is implemented as a convolutional neural network. The initial convolution layers of the network extract features from the utterance while fully connected layers are later added to predict the output probabilities and coordinates. Our network architecture is inspired by PyTorch's \cite{paszke2017automatic} implementation of the VGG19 \cite{simonyan2014very} architecture\footnote{\tiny{\texttt{https://github.com/pytorch/vision/blob/master/torchvision/models/vgg.py}}}, and is presented in Section \ref{sec:experiments}. 

The training set is composed of examples, where each example is an event that is composed of the tuple $(\sx, k, t^k_\mathrm{start}, t^k_\mathrm{end})$. We denote by  $\mathbbm{1}^{k}_{i}$ the indicator that is 1 if the keyword $k$ was uttered within the cell $c_i$, and 0 otherwise. Formally, 
\[
\mathbbm{1}^{k}_{i} = \left\{ \begin{array}{ll}
1 & t_{c_i} \le t^k  \le t_{c_{i+1}}-1 \\
0 & \mathrm{otherwise} 
\end{array} \right.~.
\]
where $t^k$ is defined as the center of event $k$.

When we would like to indicate that the keyword is not in the cell we will use the notation $(1-\mathbbm{1}^{k}_{i})$.

The model's loss function is defined as a sum over several terms, each of which took into consideration a different aspect of the model, as a follows:
\begin{multline}
\label{eq:loss} \nonumber
 \bar{\ell}(\sx, k, t^k_\mathrm{start}, t^k_\mathrm{end}) = \lambda_{1}  
 \sum_{i = 1}^{C} \sum_{j = 1}^{B}  
 \mathbbm{1}^{k}_{i} \left(t_j - t'_j\right)^2  \\
+ \lambda_{2} \sum_{i = 1}^{C} \sum_{j = 1}^{B}  
\mathbbm{1}^{k}_{i} \left( \sqrt{\Delta t_j\vphantom{\Delta t'_j}} - \sqrt{\Delta t'_j} \right)^2 \\
+ \sum_{i = 1}^{C} \sum_{j = 1}^{B} \mathbbm{1}^{k}_{i} \Big(1 - \pbij\Big)^2 \\
  + \lambda_{3} \sum_{i = 1}^{C} \sum_{j = 1}^{B} 
 \left( 1- \mathbbm{1}^{k}_{i} \right)
 \Big(0 - \pbij\Big)^2   \\
 + \sum_{i = 1}^{C} \sum_{k \in \mathcal{L}} \mathbbm{1}^{k}_{i} \Big(1-\pci(k)\Big)^2.
\end{multline}

We would like to note that our system is inspired by the first version of YOLO \cite{redmon2016you}. Further research on YOLO has been conducted in \cite{redmon2017yolo9000} and \cite{redmon2018yolov3}. It seems, however, that most expansions made to their algorithm are irrelevant for our domain. In  \cite{redmon2017yolo9000} the authors' main contributions are the addition of \textit{anchor boxes}, which defines constraints on the shapes of the bounding boxes. This lead to specifying a separate class probability value for every bounding box. This is relevant when dealing with a multidimensional domain, and is less relevant for speech. In their paper, they additionally suggest the usage of a fully convolutional network, i.e. replacing the fully connected layers with convolutions. We found that this yielded inferior results for our dataset. In \cite{redmon2018yolov3}, the main development  was the shift from multiclass classification to multilabel classification. This changed the loss function from using regression to using cross entropy instead. This too is irrelevant for our domain.



\section{Experiments}
\label{sec:experiments}

We used data from the LibriSpeech corpus \cite{panayotov2015librispeech}, which was derived from read audio books. The training set consisted of 960 hours of speech. This corpus had two test sets: \textit{test\_clean} and \textit{test\_other}, which summed up to 5 hours of speech. The first set was composed of high quality utterances and the second set was composed of lower quality utterances. The audio files were aligned to their given transcriptions using the Montreal Forced Aligner (MFA) \cite{mcauliffe2017montreal}. We extracted the Short-Time Fourier Transform (STFT) as features to the sound files using the \texttt{librosa} package \cite{librosa}. These features were computed on a 20 ms window, with a shift of 10 ms. 

A target event of the input speech signal can be defined as any discrete part of an utterance that is discernible to a human annotator. Hence, events could be defined as a set of words, phrases, phones, etc. It was assumed that only events from the selected lexicon $\mathcal{L}$ are available during training time.

\begin{figure}[t]
 \centering
 \includegraphics[width=\linewidth]{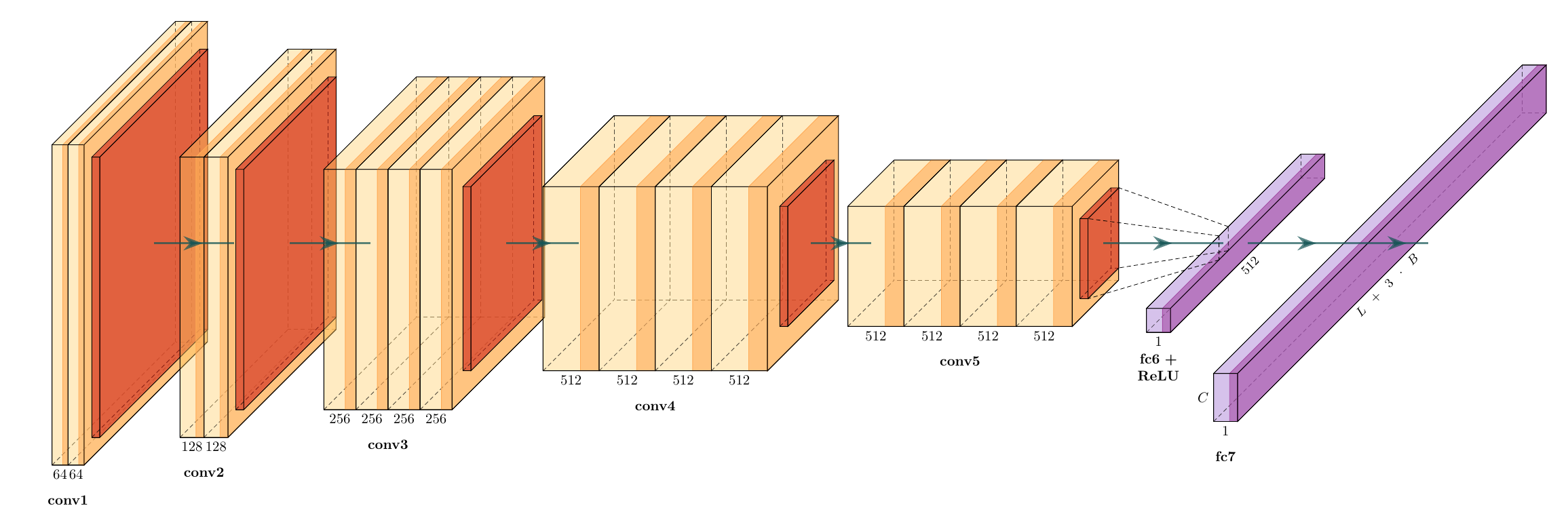}
 \caption{The detection network has 16 convolutional layers followed by 2 fully connected layers. Every convolutional layer is followed by BatchNorm and ReLU. We pretrained the convolutional layers on the Google Command classification task and then replace the final layer for detection and localization.}
 \label{fig:vgg19}
\end{figure}

We used a convolutional neural network that is similar to the VGG19 architecture.  It had 16 convolutional layers and 2 fully connected layers, and the final layer predicted both class probabilities and timing boxes' coordinates. We denote this architecture as $VGG19^*$. The model is described in detail in Figure \ref{fig:vgg19}. For comparison, we also implemented a version of the  VGG11 model (denoted by $VGG11^*$), which had less convolutional layers. Both models were trained using Adam \cite{kingma2014adam} and a learning rate of $10^{-3}$. We pretrained our convolutional network using the Google Command dataset \cite{warden2018speech} for  $L= 30$ . We later replaced the last linear layer in order to perform prediction on a different number of events, and further trained the network. The size of this new final layer is $C \times (L + 3 \cdot B)$.



We divided our experiments into two parts: in the first, we evaluated SpeechYOLO's capability to correctly predict and localize words within an utterance, and compared its performance to other similar systems. Then, we evaluated SpeechYOLO for the keyword spotting task on various domains.

\subsection{Word prediction and localization}

In this subsection, we evaluated the system's capability to learn word detection and localization. We defined the target events to be the 1000 most common words in the training set ($L=1000$). It turned out that the average utterance time of a single word in our corpus was approximately 0.2 seconds. To assure that the timing cells properly covered the span of the speech signal, we chose to use $C=6$ timing cells per utterance of $T=1$ sec. We arbitrarily set the number of timing boxes per cell to be $B=2$. 

We chose the value of the threshold $\theta$ that maximizes the F1 score, which is defined as the harmonic mean of the precision and recall. We evaluated the model's detection capabilities using Precision and Recall. Results are presented in Table \ref{tab:yolo_eval}. It seems that the proposed system was able to correctly detect most of the words, with $VGG19^*$ outperforming $VGG11^*$, due to its size and enhanced expressive abilities. 



\begin{table}[th]
  \caption{SpeechYOLO evaluations with two architectures on both of LibriSpeech's test sets. The threshold value that maximized the F1 score was chosen ($\theta = 0.4$).}
  \label{tab:yolo_eval}
  \centering
  \resizebox{\columnwidth}{!}{\begin{tabular}{@{\extracolsep{4pt}}llcccccc}
  \toprule
    & &  \textbf{precision} & \textbf{recall} & \textbf{F1}  \\
    \midrule
    \textit{test\_clean} & $VGG11^*$      & 0.743 & 0.637   & 0.686      \\
     & $VGG19^*$      & \textbf{0.836} & \textbf{ 0.779}   & \textbf{0.807}   \\
     \hline
    \textit{test\_other} & $VGG11^*$      & 0.547 & 0.456 & 0.498      \\
    & $VGG19^*$     & 0.697 & 0.553 & 0.617 \\
    \bottomrule
  \end{tabular}}
\end{table}


\subsubsection{Prediction and Localization}
Due to the uniqueness of our system's aim to both classify and localize words, it is challenging to find an equivalent algorithm to justly compare with. Most other algorithms focus on either one of the tasks, but not on both. The system of Palaz, Synnaeve and Collobert \cite{palaz2016jointly} was developed for weakly-supervised word recognition; that is, its aim is to perform word classification and find word position, while training with a BoW supervision. As in \cite{kamper2017visually}, we refer to this system as PSC.

PSC receives the Mel Filterbanks coefficients as input features. Their architecture is composed of 10 convolutional layers. The final convolution has 1000 output filters for every time span, with every filter corresponding to a word $k$ in the lexicon $\mathcal{L}$. The idea is that the score for word $k$ would be highest in the time span it occurred in. The system is trained using SGD with a learning rate of $10^{-5}$.

We compared SpeechYOLO's prediction and localization abilities to PSC's, as shown in Table \ref{tab:yolo_vs_psc}. We calculated the F1 measure as before. The \textit{Actual} accuracy measure was calculated as described in \cite{palaz2016jointly}, and measures localization as well as prediction. For PSC, the \textit{Actual} accuracy was calculated as follows: word detection was performed by thresholding the probability of a word being present in the sequence. For every word $k$ that passed the chosen threshold, we chose the frame  in which it received the highest score. We then assessed if this frame was located within the range of $k$ stated by the ground truth alignment. The closest equivalent of this measure for our model was to choose this frame to be the center of the predicted timing box. This value was in turn compared to the ground truth alignment. As before, the threshold $\theta$ was chosen to maximize the F1 measure. SpeechYOLO clearly outperformed PSC for both the F1 score and the Actual accuracy measure.

To assess the strength of SpeechYOLO's localization ability, we calculated both systems' average intersection over union (IOU) value with the ground truth alignments. While SpeechYOLO's IOU value clearly outperformed that of PSC, one must remember that PSC was not trained with aligned data.

\begin{table}[th]
  \caption{Comparing SpeechYOLO and PSC \cite{palaz2016jointly}'s evaluations of the F1 score, Actual accuracy and average IOU value. The threshold value that maximized the F1 score was chosen for every algorithm separately ($\theta = 0.4$ ). }
  \label{tab:yolo_vs_psc}
  \centering
  \begin{tabular}{l c c c}
    \toprule
   \multicolumn{1}{c}{} & \multicolumn{1}{c}{\textbf{F1}} & 
   \multicolumn{1}{c}{\textbf{Actual}} & \multicolumn{1}{c}{\textbf{IOU}} \\
    \midrule
    SpeechYOLO  & \textbf{0.807} & \textbf{0.774}   & \textbf{0.843} \\
    PSC                     & 0.767  & 0.692  & 0.3   \\
    \bottomrule
  \end{tabular}
\end{table}

We further checked the quality of SpeechYOLO's localization capability. To do so, we compared SpeechYOLO with MFA, after both had been trained on LibriSpeech. We tested them on the training set of the TIMIT corpus. TIMIT is a corpus of read speech, and presents a different linguistic context compared to LibriSpeech. The IOU measure was used to compare both algorithm's output alignments with TIMIT's given word alignments for the 1000 most common words in the LibriSpeech training set. In order to predict SpeechYOLO's IOU values, it was assumed that its predictions were perfect. This was due to the fact that SpeechYOLO does not receive transcriptions as an input, and because our goal was to asses the localization task alone.  The IOU of SpeechYOLO on TIMIT was 0.673, while MFA achieved 0.827.
 
The forced aligner, MFA, performs its alignments using a complete transcription of the words uttered in a speech signal. On the other hand, SpeechYOLO receives no information about the words uttered. Hence, given MFA's extended knowledge, we considered its localization ability as an ``upper bound'' to ours. Therefore, we found that SpeechYOLO's IOU value, while lower than MFA's, were sufficiently high.


Additionally, an aligner could naturally go wrong if there are incorrect or missing words in its transcription, or alternatively if the audio signal contains long silences or untranscripted noises between words (e.g. a laugh or a cough) \cite{chodroff2018corpus}. It should be noted that given SpeechYOLO's lack of knowledge about the transcription, these problems do not affect it. 

\subsubsection{Robustness to noise}
\begin{figure}
\centering
\begin{subfigure}{\columnwidth}
  \centering
  \includegraphics[width=.7\columnwidth]{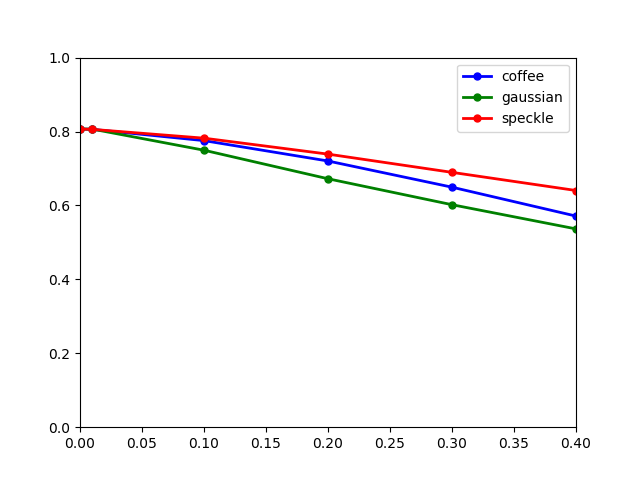}
  \caption{F1 score}
  \label{fig:sub1}
\end{subfigure}%
\newline
\vspace{-.1cm}
\begin{subfigure}{\columnwidth}
  \centering
  \includegraphics[width=.7\columnwidth]{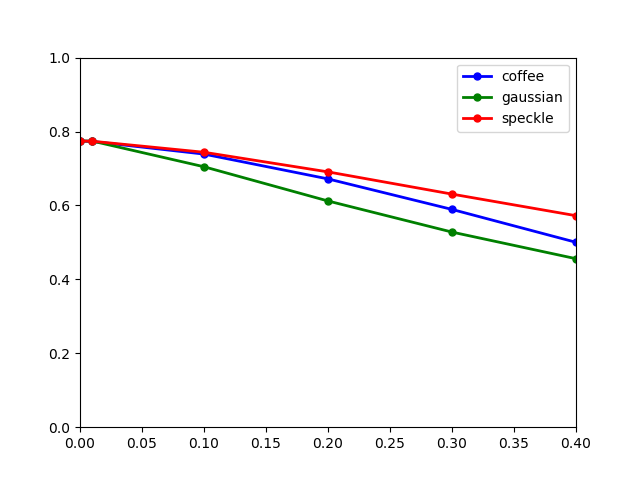}
  \caption{ Actual accuracy}
  \label{fig:sub2}
\end{subfigure}
\caption{SpeechYOLO's performances when injecting background noise. The y-axis is the measure and the x-axis is the strength of the noise added ($\alpha$).}
\label{fig:noise}
\vspace{-.5cm}
\end{figure}

We further demonstrated SpeechYOLO's robustness by artificially adding background noise to LibriSpeech's audio files with a relative amplitude $\alpha$.  
We injected 3 types of background noises: a coffee shop, gaussian noise, and speckle. In Figure \ref{fig:noise} we show SpeechYOLO's F1 score and \textit{Actual} accuracy measures when increasing the $\alpha$ variable, thus intensifying the injected noise. It is apparent that minor amounts of noise do not degrade SpeechYOLO's performances. Note that SpeechYOLO was able to deal even with higher $\alpha$ values, although it yielded somewhat reduced results.

\subsection{Keyword spotting}

In this part, we evaluate SpeechYOLO on a real-world application: keyword spotting. For evaluation, we use the F1 metric, and the Maximum Term Weight Value (MTWV) metric \cite{fiscus2007results}.  MTWV is defined as one minus the weighted sum of the probabilities of miss and false alarm. 

\subsubsection{LibriSpeech Corpus}
We compare SpeechYOLO's keyword spotting capabilities with those of the PSC system. In their work, they use a set of keywords that is a subset of the 1000 words used previously for prediction and localization. The chosen keywords are in Table 2 of \cite{palaz2016jointly}, and are evaluated on both of LibriSpeech's test sets. A comparison of our results are presented in Table \ref{tab:yolo_vs_psc_kws}. Here too SpeechYOLO's results outperformed those of PSC.


\begin{table}[th!]
  \caption{ MTWV values for SpeechYOLO and PSC on the keyword spotting task, evaluated on both of LibriSpeech's test sets.}
  \label{tab:yolo_vs_psc_kws}
  \centering
  \begin{tabular}{l cc}
    \toprule
   \multicolumn{1}{c}{}  & \multicolumn{1}{c}{\textbf{SpeechYOLO}}  & \multicolumn{1}{c}{\textbf{PSC}}\\
    \midrule
    \textit{test\_clean}          & \textbf{0.74} \ignore{($\theta = 0.4$)} &   0.72 \ignore{($\theta = 0.3$)}   \\
    \textit{test\_other}         & \textbf{0.38} \ignore{($\theta = 0.6$)}  &  0.27 \ignore{($\theta = 0.2$)} \\
    \bottomrule
  \end{tabular}
\end{table}

\subsubsection{Spontaneous speech corpus}
We now present the results of SpeechYOLO for keyword spotting with spontaneous speech. This is relevant for mobile applications, where a device is activated by a voice command like `OK Google'' or ``Hey Siri''. To simulate this task, we use a corpus taken from a daily TV show \textit{Good evening with Guy Pines}\footnote{\tiny{\texttt{https://www.timesofisrael.com/topic/good-evening-with-guy-pines/}}}. This corpus, which we will call ``Hi Guy'', consists of spontaneous and noisy recordings. In each recording, a celebrity is prompted to utter the phrase \textit{Hi Guy!} These recordings vary greatly in terms of their environment and the speakers within them are highly diverse. 

The corpus consists of 880 examples, out of which 445 contain the chosen keyword. We chose the phrase ``Hi Guy'' to be the keyword that our system searches for. The input length is 3 seconds. The system achieves an \textit{Actual} accuracy of 0.624, and an F1 score of 0.755 (precision: 0.748, recall: 0.761). We find these results to be surprisingly satisfying due to the small size of the dataset and due to the diversity found in the corpus: the audio files are at times extremely noisy, the pronunciation of the speakers vary, and the keyword is sometimes sung instead of being spoken. 


  

\vspace{-.2cm}

\section{Conclusions}
\label{seq:conclusions}

In this work, we introduce the concept of treating parts of audio signals as objects. We propose the SpeechYOLO algorithm for object detection and localization, and evaluate its performances for both of these tasks. We further show its use for the keyword spotting task. Future work includes expanding its keyword spotting ability for other speech parts. We would also like to extend SpeechYOLO to detect words it has not been trained on.

\section{Acknowledgements}

We would like to thank \textit{Hi Guy! Guy Pines Communications Ltd.} for providing their data. T. S. Fuchs is sponsored by the Malag scholarship for outstanding doctoral students in the high tech professions.

\bibliographystyle{IEEEtran}
\bibliography{mybib}

\end{document}